\documentclass[10pt,a4paper,prd,tightenlines,nofootinbib,showpacs,showkeys,twocolomn, onecolomn, superscriptaddress, notitlepage,reprint]{revtex4-1}
\usepackage{bm}

\usepackage[pdftex]{graphics}
\usepackage{rotating}
\usepackage{epsfig}
\usepackage[usenames]{color}
\usepackage{epstopdf}
\usepackage{mathtext}
\usepackage{amsmath}

\begin{document}
\title{How disorder originates and grows inside order}

\author{\firstname{S.\,E.} \surname{Kurushina}}
\affiliation{Physics Department, Samara National Research University named after S.P. Korolyov, 34, Moskovskoye shosse, Samara, 443086, Russian Federation}
\author{ \firstname{E.\,A.} \surname{Shapovalova}}
\affiliation{Physics Department, Samara National Research University named after S.P. Korolyov, 34, Moskovskoye shosse, Samara, 443086, Russian Federation}

\begin{abstract}
We present a study of disorder origination and growth inside an ordered phase processes induced by the presence of multiplicative noise within mean-field approximation. Our research is based on the study of solutions of the nonlinear self-consistent Fokker-Planck equation for a stochastic spatially extended model of a chemical reaction. We carried out numerical simulation of the probability distribution density dynamics and statistical characteristics of the system under study for varying noise intensity values and system parameter values corresponding to a spatially inhomogeneous ordered state in a deterministic case. Physical interpretation of the results obtained that determines the scenario of noise-induced order-disorder transition is given. Mean-field results are compared with numerical simulations of the evolution of the model under study.

We find that beginning from some value of external noise intensity the ``embryo" of disorder appears inside the ordered phase. Its lifetime is finite, and it increases with growth of noise intensity. At some second noise intensity value the ordered and disordered phases begin to alternate repeatedly and almost periodically. The frequency of intermittency grows with the increasing of noise intensity. Ordered and disordered phase intermittency affects the process of spatial pattern formation as a consequent change of spatial inhomogeneity configurations.
\end{abstract}



\maketitle

\twocolumngrid{

\section{Introduction}

The mean-field concept is widely used in physics of macro- and microworld to study the dynamics of complex stochastic systems consisting of a large number of individual components interacting with each other. Both many-particle systems with internal interaction and spatially extended systems belong to this class of systems. Various approximations and averaging procedures are proposed within this concept, nevertheless, the main idea of the mean-field theory (MFT) implies the replacement of all interactions with any one component by average interaction.

The Ising model is one of the examples of many-particle systems with internal interaction. In Refs.  \cite{Devir1,RiegoBerg2,VelStarBill3} dynamic phase transitions are studied in a mixed spin-3/2 and spin-5/2 Ising system with a crystal-field interaction under a time-varying magnetic field \cite{Devir1}  and kinetic Ising model in systems with surfaces \cite{RiegoBerg2}. In Refs. \cite{Devir1,RiegoBerg2} the mean-field approximation (MFA) is used to simplify the Hamiltonian representation so as to obtain the appropriate dynamic equations. The origin of the inverse symmetry-breaking transition in ultrathin magnetic films with perpendicular anisotropy is theoretically studied in Ref. \cite{VelStarBill3} by using a dipolar frustrated Ising model in a two-dimensional square lattice. Here MFA assumes that magnetization is equal to a mean value of the Ising spin, and applied to write the free energy of the model.

A neural network consisting of interacting neurons is another example. The nonequilibrium properties of a neural network which models the dynamics of the neocortex are considered in Ref. \cite{WilMooBeg4}. An analytical MFA in the form of an autonomous nonlinear discrete dynamical map of first order and dimension given by the integer-valued refractory period is developed. The network is simulated by cellular automatons. The cellular automaton rules for the cortical branching model introduced by the authors are approximated by a Markovian stochastic process. In Ref. \cite{WilMooBeg4}  at a particular iteration of the mean field the probability that some node is in a particular state is equivalent to the fraction of all nodes in this state. In a typical case, MFA means that a representative node and its local neighborhood of interaction are used to approximate the behavior of the network as a whole.

The so called ``networks of networks" are one more example. They consist of complex interconnected systems often possessing inconsistent topologies which themselves may consist of agents distributed in different interacting subnetworks. In Ref. \cite{ZhuFu38}  a scenario of epidemic spreading in such interconnected networks is considered. The heterogeneous mean-field approximation (HMFA) is used to describe the epidemic evolution. In Ref. \cite{WuChen39}  a susceptible-infected-susceptible model with effective contacts in networks where each susceptible node is infected with a certain probability only for effective contacts is studied. By using the one-vertex HMFA and the pair HMFA, the conditions for epidemic outbreak by the degree of network uncorrelatedness are obtained.

In Ref. \cite{TakYas5} two different types of naive mean-field approximation (NMFA) for Gaussian restricted Boltzmann machine are derived. NMFA are obtained by minimizing the Kullback-Leibler divergence with respect to two test distributions.

In Ref. \cite{Chavanis6} the dynamics and thermodynamics of the Brownian Mean Field (BMF) model which contain Brownian particles moving on a circle and interacting via a cosine potential, are discussed. In Ref.\cite{Cebers7}  kinetic equations are derived within the MFA for dielectric particles coupled due to the flow of translational and rotational motion. The studies presented in Ref.\cite{KudrDrchal8} demonstrate that a correct relation of Curie temperatures of $Fe_{3}Al/Fe_{3}Si$ alloys can be obtained only by going beyond a simple mean-field approximation. The existence of phase transitions in an intermediate model between a class of probabilistic cellular automata, called majority voters, and their corresponding mean-field models is shown in Ref.\cite{Bricmont9}.

MFA has found wide application for the study of quantum phenomena in Refs.\cite{Ayik10,Horv11,Graefe12,Akerlund13,Sowi14,Bighin15,Hayami16,Dixit17,Leeuw18,Nasu19,Rosati20,Serreau21,Vermersch22,Yilmaz23,YilmazAyik24}. Among other things, in Ref. \cite{Ayik10} a microscopic stochastic approach is proposed to improve the description of nuclear dynamics beyond the MFA at low energies. In Ref.\cite{Graefe12} MFA based on a coherent state approximation is derived to investigate an N-particle Bose-Hubbard dimer with an additional effective decay term in one of the sites. The zero temperature properties of the generalized Bose-Hubbard model including three-body interactions are studied in Ref.\cite{Sowi14} using the so-called perturbative mean-field method. MFA is used in Ref.\cite{Serreau21} to derive the most general evolution equations describing in-medium (anti)neutrino propagation; in Ref. \cite{Vermersch22}  it is used to study the evolution of a system of interacting ultracold bosons that exhibit nonlinear chaotic behavior in the limit of a very large number of particles. In Ref. \cite{Reich25} optimal quantum control of the dynamics of trapped Bose-Einstein condensates is studied. The condensate dynamics is described by the Gross-Pitaevskii equation in the MFA. MFA is also used to study the dynamics of an open two-mode Bose-Hubbard system subject to phase noise and particle dissipation in Ref. \cite{Trimborn26} and to consider a Bose-Einstein-condensed cloud of atoms which rotate in a toroidal or annular potential in Ref. \cite{Roussou27}. The method is applied to study chiral phase transitions in Refs. \cite{Morita29,Skokov30}.

An important feature of the MFA is the possibility of its application to study dynamical systems described by systems of differential equations. The Ref. \cite{Vannucchi32} studies the model of forming public opinion in a community united through a general network. The master equation describing the time evolution of opinions is presented and solved in MFA. To perform the MFA authors assume no correlation of opinion probabilities, opinion is homogeneous over the network, and all the local variables may be replaced by their mean values all over the network. In Ref. \cite{Aguiar34} the MFA is used for a simple spatial host-pathogen model to demonstrate its interesting evolutionary properties.

Models for surface growth with a wetting and a roughening transition are studied using simple and pair MFA in Ref. \cite{Barato33}. In simple MFA all the correlations are neglected. In the pair MFA approximation for the three-site probability distribution which contains two-site and one-site probability distributions is used. Simple MFA makes it possible to predict the roughening transition and the correct growth exponents in a region of the phase diagram. Pair MFA correctly predicts a growing interaction with constant velocity in the moving phase.

In some cases the MFA means the volume averaging simply, for example in Ref. \cite{Campanelli31}.

In Ref. \cite{Vrettas37} dynamical systems that can be modeled by ordinary stochastic differential equations are considered. Gaussian variational MFA is introduced for studying them. This approach allows one to express the variational free energy as a functional of the marginal moments of the approximating Gaussian process. In Ref. \cite{Franovi35} a system of two nonlinear stochastic differential equations with delay wherein stochastic increments are defined as the increments of independent Wiener processes is considered. Two types of approximations are presented: the quasi-independence approximation (random variables are approximately equal to their expectations at each time moment and for a sufficiently large number of equations); Gaussian approximation (for most time instances small random increments can be computed with sufficiently good accuracy assuming that the random variables are normally distributed around their expectations for each time moment).

In Ref. \cite{Kolmakov36} the diffusive dynamics of exciton polaritons in an optical microcavity with an embedded molybdenum disulfide monolayer is considered. The relevant range of parameters at which room-temperature superfluidity can be observed is determined experimentally. The MFA is used in numerical simulations.

MFA  is also modified to study noise-induced phenomena in spatially extended systems. In Ref.\cite{Lindnera40} the behavior of theoretical models of excitable systems driven by Gaussian white noise is reviewed. Noise-induced phase transitions, noise-induced oscillations, stochastic resonance, stochastic synchronization, and other phenomena arising in the presence of noise in the FitzHugh-Nagumo and other models are studied. In Ref. \cite{Carrillo41} models described by stochastic partial differential equations with multiplicative noise are studied. Linear instability at the transition point is absent in the models. These models exhibit an ordering transition independently of interpretation of noise in the sense of Stratonovich or Ito, although the locations of the critical points are different. In Ref. \cite{Zaikin42} it is shown that a nonequilibrium first-order phase transition can be induced by additive noise in a nonlinear lattice of overdamped oscillators with both additive and multiplicative noise terms. In Ref. \cite{Landa44} it is shown, that in nonlinear chains additive noise significantly shifts the boundaries of the phase transition induced by multiplicative noise. In Ref. \cite{Carrillo43} two stochastic reaction-diffusion models with nonlinear reaction terms in the form of fifth-degree polynomials are studied. The models exhibit first and second-order nonequilabrium phase transitions induced by external nonlinear multiplicative noise. It is shown, that these transitions are caused by nonlinear instability of the homogeneous phase. In Ref. \cite{Buceta45} the behavior of a random Ginzburg-Landau model, where quenched dichotomous noise affects the control parameter is analyzed. It is shown that as the coupling increases the system exhibits a disorder-order-disorder transition whereas an increase of the quenched noise intensity gives rise to an order-disorder-order transition. These transitions are both reentrant second-order phase transitions. The phase diagram of the model has a saddle-point structure. In Refs. \cite{Lindnera40,Carrillo41,Zaikin42,Landa44,Carrillo43,Buceta45}  MFA   consists in assuming that the nearest-neighbor conditional average values of exact field in Langevin or Fokker-Planck equations are replaced by the general value of the average field. In Ref. \cite{Iba46} a modified MFA is proposed to study the phase transitions in conserved dynamics systems and the phenomenon of noise-induced phase separation is studied. In Ref. \cite{Kurushina47} the MFA for multicomponent stochastic spatially extended systems is developed and applied to study spatial pattern formation and disruption under external noise.

The review above demonstrates the manysidedness and flexibility of the mean-field approach, and proves its efficiency for studying a wide range of noise-induced phenomena and dynamics of various types of stochastic systems.

The purpose of this paper is to study the processes of disorder origination and growth induced by the presence of external noise inside an ordered phase within the mean-field approximation.

The rest of the paper is organized as follows. In Sec. II we review the main points of the MFA in its application to multi-component stochastic reaction-diffusion systems. We consider a chemical reaction model --- a stochastic spatially extended brusselator as a model system to study these processes. In Sec. III the model under study and nonlinear self-consistent Fokker-Planck equation (NSCFPE) for it are presented. The various types of solutions of this equation are described. In Sec. IV the noise intensity values are defined corresponding to the types of NSCFPE solutions in which a bimodality of the probability distribution density occurs that means the ordered state begins to disrupt. The dynamics of probability distribution density and statistical characteristics of the first and second orders of product concentrations in the specified region are presented. Physical interpretation of the results obtained is given. In Sec. V the results of numerical simulation of the evolution of the systems under study are presented. The results confirm the theoretical conclusions of Sec. IV. Finally, some conclusions are reported in Sec. VI.

\section{MFA for multicomponent stochastic reaction-diffusion systems}

We now review the main points of  MFA  in its application to multicomponent stochastic reaction-diffusion systems, in order to explain the two-dimensional NSCFPE, on the basis of the solutions of which the dynamics of the chemical reaction model under study is analyzed and that is presented in the next section.

Let us consider a generalized model - a system of stochastic reaction-diffusion equations:

\begin{equation}
\label{eq1}
\begin{array}{l}
\frac{\partial x_{i}} {\partial t} =  f_{i} (x_{1} ,...,x_{n} ) + g_{i}(x_{1} ,...,x_{n} )\xi _{i} (\mathbf{r},t)  \\
\\
  \  \  \  \  \  \  \  \  \  + \eta _{i} (\mathbf{r},t)+ D_{i} \nabla ^{2}x_{i}, \, i = 1,...,n.\\
\end{array}
\end{equation}

\noindent
In  Eq. (\ref{eq1}) $x_{i}$  are the functions defining the system state, containing $n$ components,  $f_{i} (x_{1}
,...,x_{n} ),\, g_{i} (x_{1} ,...,x_{n} )$  are nonlinear functions defining the interaction and evolution of component $x_{i}$  in space and in time, $D_{i}$  are diffusion coefficients of components. The additive random Gaussian fields $\eta _{i} (\mathbf{r},t)$  with zero mean and correlation functions $K{\left[ {\eta _{i} (\mathbf{r},t),\eta
_{{i}'} (\mathbf{{r}'},{t}')} \right]} = 2\zeta _{i} \delta (\mathbf{r} - \mathbf{{r}'})\delta (t - {t}')\delta _{i{i}'} $ simulate the internal white noises with intensities $\zeta _{i} $. The multiplicative random Gaussian homogeneous and spatially isotropic fields  $\xi _{i} (\mathbf{r},t)$  with zero mean and correlation functions $K{\left[ {\xi _{i} (\mathbf{r},t),\xi _{{i}'} (\mathbf{{r}'},{t}')} \right]} = 2\theta _{i} \Phi _{i} ({\left| {\mathbf{r} - \mathbf{{r}'}} \right|})\delta (t - {t}')\delta _{i{i}'} $  simulate external noises.  $\Phi_{i} ({\left| {\mathbf{r} - \mathbf{{r}'}} \right|})$  are spatial correlation functions of the external noises and $\theta _{i} $   are their intensities. Further, for definiteness, we use exponential spatial correlation functions $\Phi _{i} ({\left| {\mathbf{r} - \mathbf{{r}'}} \right|}) = \exp [ - k_{fi} ({\left| {\mathbf{r} - \mathbf{{r}'}} \right|})]$, where the correlation lengths $r_{fi} $  of the noises are defined as $r_{fi} = 1 / k_{fi} $. Hereafter the notation $K[F_{1} ,F_{2} ]$  defined by the equality $K[F_{1},F_{2} ] = < F_{1} F_{2} > - < F_{1} > < F_{2} > $  is used for the correlation function.

We carry out the discretization of continuous $d$-dimensional space of system (\ref{eq1})  and obtain a regular $d$-dimensional lattice with mesh size $\Delta r=1$  and points the locations of which are characterized by vectors $\mathbf{r}_{l} ,\,l = 1,...,p.$   We assume that the interaction takes place only between the nearest neighbors, then the Laplace operator can be approximated by a finite-difference expression with a second-order difference. As a result of the discretization the system (\ref{eq1})  is replaced by the system of $n\times p$  ordinary differential equations:

\begin{equation}
\label{eq2}
\begin{array}{l}
{\frac{{dx_{il}}} {{dt}}} = F_{il} (t),\,i = 1,...,n;\,l = 1,...,p,\\
\\
F_{il} (t) = f_{il} + g_{il} \xi _{il} (t) + \eta _{il} (t) + {\frac{D_{i}}{2d}}{\sum\limits_{{l}'} {\Lambda _{l{l}'}}}  x_{i{l}'}.\\
\end{array}
\end{equation}

\noindent
In Eqs. (\ref{eq2}) the following notations are introduced: $f_{il} = f_{i} (x_{1l}
,...,x_{nl} ),\,\,g_{il} = g_{i} (x_{1l} ,...,x_{nl} ).$
${\sum\nolimits_{{l}'} {\Lambda _{l{l}'}}}  $ is the discrete analog of the
Laplace operator \cite{Iba46}: ${\sum\nolimits_{{l}'} {\Lambda _{l{l}'}}}   =
{\sum\nolimits_{{l}'} {(\delta _{nn(l),{l}'} - 2d\delta _{l,{l}'} )}} $,
where $nn(l)$ is a set of indexes of all points being the nearest neighbors of the point with index  $l$. The discrete noises $\eta _{il} (t),\,\,\xi _{il} (t)$ have the correlation functions:

\noindent
$ K{\left[ {\eta _{il} (t),\eta _{{i}'{l}'} ({t}')} \right]} = 2\zeta _{i}
\delta _{l{l}'} \delta (t - {t}')\delta_{i{i}'}$   and

\noindent
$K{\left[ {\xi _{il} (t),\xi _{{i}'{l}'} ({t}')} \right]} =
2\theta _{i} \Phi _{i,\vert l - {l}'\vert}  \delta (t - {t}')\delta _{i{i}'}.$

\noindent
The continuous delta function $\delta (\mathbf{r} - \mathbf{{r}'})$  here is replaced by the Kronecker delta-symbol
$\delta _{l{l}'},$   $\Phi _{i,\vert l - {l}'\vert}  $ is the discrete analog of function  $\Phi _{i} ({\left| {\mathbf{r} -
\mathbf{{r}'}} \right|})$.

The Fokker - Planck equation in the Stratonovich interpretation \cite{Strat49} corresponding to  Eqs. (\ref{eq2}) for multivariate probability density $\tilde {w}(x_{11} ,...,x_{1l} ,...,x_{1p} ,...,x_{n1} ,...,x_{nl}
,...,x_{np} ;t) = \tilde {w}$ for all lattice points  has the form:
}
\onecolumngrid{

\begin{equation}
\label{eq3}
\frac{\partial \tilde {w}}{\partial t} =
- {\sum\limits_{i = 1}^{n} {{\sum\limits_{{l}' = 1}^{p} {{\frac{{\partial
}}{{\partial x_{i{l}'}}} }}}} } {\left[ {f_{i{l}'} + {\frac{D_{i}}{2d}}\left( {{\sum\limits_{m = nn({l}')} {x_{im}}}   -
2dx_{i{l}'}}  \right) -  \\
{\sum\limits_{m = {l}',nn({l}')} {\left( {\zeta _{i}
{\frac{{\partial}} {{\partial x_{im}}} } - \theta _{i} g_{i{l}'} \Phi
_{i,\vert {l}' - m\vert}  {\frac{{\partial}} {{\partial x_{im}}} }g_{im}}
\right)}}}  \right]}\tilde {w}.
\end{equation}

We choose one point with index $ l$. In order to obtain multivariate probability density $w(x_{1l} ,...,x_{il} ,...,x_{nl} ;t) = w$  for a single point, it is necessary to integrate $\tilde {w}$   over all the variables except  $x_{1l} ,...,x_{il}
,...,x_{nl} $. Using the properties of the probability density and definition of the conditional probability, finally, we get for a multivariate probability density at one point $ l$:

\begin{equation}
\label{eq4}
{\frac{{\partial w}}{{\partial t}}} = - {\sum\limits_{i = 1}^{n} {{\frac{{\partial}} {{\partial x_{il}}} }}
}{\left\{ {f_{il} + {\frac{{D_{i}}} {{2d}}}\left[
{{\sum\limits_{m = nn(l)} {E(x_{im} \vert x_{1l} ,...,x_{il} ,...,x_{nl}
;t)}}  - 2dx_{il}}  \right] - \zeta _{i} {\frac{{\partial}} {{\partial
x_{il}}} } - \theta _{i} \Phi _{i,0} g_{il} {\frac{{\partial}} {{\partial
x_{il}}} }g_{il}}  \right\}}w
\end{equation}

\noindent
Here $E(x_{im} \vert x_{1l} ,...,x_{il} ,...,x_{nl};t)$ are point l nearest-neighbor conditional averages.

As $x_{il} $  are related by the Eqs.(\ref{eq2}) the MFA here consists in the assumption that the conditional averages in Eq. (\ref{eq4})  is replaced by

\begin{equation}
\label{eq5}
E(x_{im} \vert x_{1l} ,...,x_{il} ,...,x_{nl} ;t) = E({\left. {x_{il}}
\right|}x_{1l} ,...,x_{i - 1l} ,x_{i + 1l} ,...,x_{nl} ;t),
\end{equation}

\noindent
where

\[
\begin{array}{l}
 E({\left. {x_{il}}  \right|}x_{1l} ,...,x_{i - 1l} ,x_{i + 1l} ,...,x_{nl}
;t) = {\int\limits_{ - \infty} ^{ + \infty}  {x_{il}}}  w({\left. {x_{il}}
\right|}x_{1l} ,...,x_{i - 1l} ,x_{i + 1l} ,...,x_{nl} ;t)dx_{il} , \\
 w({\left. {x_{il}}  \right|}x_{1l} ,...,x_{i - 1l} ,x_{i + 1l} ,...,x_{nl}
;t) = {\frac{{w(\{x\};t)}}{{{\int\limits_{ - \infty} ^{ + \infty}  {w(x_{1l}
,...,x_{il} ,...,x_{nl} ;t)dx_{il}}} } }}. \\
 \end{array}
\]

In this approximation the exact FPE (\ref{eq4})   is transformed into an approximate equation

\begin{equation}
\label{eq6}
{\frac{{\partial w}}{{\partial t}}} =- {\sum\limits_{i = 1}^{n} {{\frac{{\partial}} {{\partial x_{il}}} }}
}{\left\{ {f_{il} + {D_{i}}\left[ {E({\left.
{x_{il}}  \right|}x_{1l} ,...,x_{i - 1l} ,x_{i + 1l} ,...,x_{nl} ;t) -
x_{il}}  \right] - \zeta _{i} {\frac{{\partial}} {{\partial x_{il}}} } -
\theta _{i} \Phi _{i,0} g_{il} {\frac{{\partial}} {{\partial x_{il}
}}}g_{il}}  \right\} }w.
\end{equation}

}

\twocolumngrid{
\noindent
Here, the index $ l$  is dropped for simplicity.

Eqs. (\ref{eq5},\ref{eq6})  form a self-consistent system, the numerical solution of which can be obtained using Ref.\cite{Kurushina47}.

\section{Two-dimensional NSCFPE for spatially extended stochastic brusselator and its solutions}

Let us consider the scheme of the very simple chemical reaction \cite{Prigogine54}:

\begin{equation}
\label{eq7}
\begin{array}{l}
A \longrightarrow X_{1},\\
2X_{1} + X_{2} \longrightarrow 3X_{1},\\
B + X_{1} \longrightarrow X_{2} + D,\\
X_{1} \longrightarrow E.\\
 \end{array}
\end{equation}

\noindent
The overall reaction is $A + B \longrightarrow D +  E.$  In Ref.\cite{Prigogine54}  it is assumed that the initial and final product concentrations are maintained constant. The scheme (\ref{eq7}) is a theoretical model of an autocatalytic reaction. It got the name of a ``Brusselator". The scheme has a trimolecular step, therefore it is physically unreal. However, despite this, it is widespread and is often used to study the self-organization phenomena due to the simplicity of mathematical analysis of its kinetic equations and the possibility to demonstrate a wide range of behavior regimes.

We assume that the concentrations of initial products are affected by external random factors and fluctuate near their average values. Neglecting the reverse reactions we write the kinetic equation of reaction (\ref{eq7}) supplementing them with terms that describe the diffusion of the intermediate components and the fluctuations of the parameters $A$ and $B$:

\begin{equation}
\label{eq8}
\begin{array}{l}
{\frac{{\partial x_{1}}} {{\partial t}}} = A + \xi _{3} (\mathbf{r},t) + x_{1} ^{2}x_{2} \\
 \  \  \  \  \  \  \  \  \  - (B + 1 + \xi _{1} (\mathbf{r},t))x_{1} + D_{1} \nabla ^{2}x_{1} ,\\
\\
{\frac{{\partial x_{2}}} {{\partial t}}} = - x_{1} ^{2}x_{2} + (B + \xi _{2}
(\mathbf{r},t))x_{1} + D_{2} \nabla ^{2}x_{2}.
 \end{array}
\end{equation}

\noindent
In Eqs.(\ref{eq8})  $x_{1}$, $x_{2}$ are concentrations of the intermediate components, $D_{1}$, $D_{2}$ are their diffusion coefficients, $A$ and $B$ are the spatiotemporal averages of the initial product concentrations. Since the concentration of the intermediate product $x_{1}$  decreases due to two different decays the equations (\ref{eq8}) contain different uncorrelated fields $\xi _{1} (\mathbf{r},t)$ and $\xi _{2} (\mathbf{r},t)$. Statistical properties of the fields $\xi _{i} (\mathbf{r},t)$   are described in Sec. II. In the deterministic case model (\ref{eq8}) has a steady state $x_{10} = A$, $ x_{20}= B/A$. When $B > B_{c}$   ($B_{c}={\left[ 1 + A{\left( D_{1}/D_{2}\right)}^{1/2} \right]}^{2} $)   instability occurs, breaking the symmetry, and the system passes to a new spatially inhomogeneous ordered state. Further, it will be the process of disrupting this kind of ordered state by the external noise that we will analyze.

}
\onecolumngrid{

Using Eq.(\ref{eq6}) for the case of $n=2$, we write two-dimensional NSCFPE for system (\ref{eq8}):

\[
{\frac{{\partial w(x_{1} ,x_{2} ,t)}}{{\partial t}}} = {\frac{{\partial
}}{{\partial x_{1}}} }{\left( {{\left\{ { - A - x_{1} ^{2}x_{2} + (B + 1 +
\theta _{1} )x_{1} - D_{1} \left[ E(x_{1} \vert x_{2} ) - x_{1} \right]} \right\}}w +
\left( \theta _{1} \Phi _{1,0} x_{1}^{2} + \theta _{3} \Phi _{3,0}\right) {\frac{{\partial w}}{{\partial x_{1}}} }}
\right)}
\]
\begin{equation}
\label{eq9}
 + {\frac{{\partial}} {{\partial x_{2}}} }{\left( {{\left\{ {x_{1} ^{2}x_{2}
- Bx_{1} - D_{2} \left[ E(x_{2} \vert x_{1} ) - x_{2} \right]} \right\}}w + \theta _{2}
\Phi _{2,0} x_{1}^{2} {\frac{{\partial w}}{{\partial x_{2}}} }} \right)},
\end{equation}

}
\twocolumngrid{
\noindent
where
\[
\begin{array}{l}
E(x_{1} \vert x_{2} ,t) = {\int\limits_{ - \infty} ^{ + \infty}  {x_{1}
w(x_{1} \vert x_{2} ,t)dx_{1}}}  ,\\
E(x_{2} \vert x_{1} ,t) = {\int\limits_{ - \infty} ^{ + \infty}  {x_{2}
w(x_{2} \vert x_{1} ,t)dx_{2}}}  ,\\
w(x_{1} \vert x_{2} ,t) = {\frac{{w(x_{1} ,x_{2} ,t)}}{{{\int\limits_{ -
\infty} ^{ + \infty}  {w(x_{1} ,x_{2} ,t)dx_{1}}} } }},\\
w(x_{2} \vert x_{1} ,t) = {\frac{{w(x_{1} ,x_{2} ,t)}}{{{\int\limits_{ -
\infty} ^{ + \infty}  {w(x_{1} ,x_{2} ,t)dx_{2}}} } }}.\\
\end{array}
\]

Primary numerical analysis showed \cite{Kurushina47} that the equation (\ref{eq9}) has three types of solutions in the parameter region indicated above. The first type of solutions, in which only unimodal probability density $ w(x_{1} ,x_{2} ,t)$  is observed until the stationary state is reached, exists at small values of noise intensity. Herewith the system (\ref{eq8}) remains in an ordered state, despite the noise. Increasing noise intensity values leads to the second type of solutions, at which the splitting of unimodal probability distribution density takes place at a certain point in time and transient bimodality arises that exists for a finite amount of time. As a result of competition one of the maxima suppresses the other and $ w(x_{1} ,x_{2} ,t)$  remains unimodal until the steady state is reached. The bimodal distribution corresponds to a disordered phase, that is, in case of this type of solution a disordered phase --- a disorder --- temporarily occurs inside the ordered phase. Further increase of the noise intensity values leads to the occurrence of a new third type of solutions --- multiple alternation of unimodal and bimodal types of probability distribution density, which corresponds to a multiple alternation of ordered and disordered phases. Herewith peculiar ``repumping" of the probability density from one maximum to another through bimodality is observed. This type of solution has practically not been studied, but it is this type that is the key to understanding of the scenario of noise-induced order-disorder transition.

\section{Disorder origination}

So, on the basis of the above we can conclude that a disordered phase first appears with the appearance of transient bimodality. It is from this point that the disruption of the order and origination of disorder begins. Therefore, at the beginning it is necessary to define the values of the problem parameters appropriate to the second type of solutions of equation (\ref{eq9}). Due to the impossibility of the analytical study of equation (\ref{eq9}) the analysis presented below is based on numerical simulations of its solutions by using the finite-difference locally one-dimensional method, the scheme of which is presented in detail in Ref.\cite{Kurushina47}.

For numerical integration (9) the following parameters, which remain constant in our calculations, are chosen: $ A=3$, $ D_{1}=1$, $D_{2}=5$, $\Phi_{1,0} = \Phi_{2,0}=1$, $\theta_{3}=0$. Under these parameters the critical value of parameter $B$ is $B_{c}=5.483$. The initial distribution is Gaussian with variances $\theta_{1}=\theta_{2}=0.01$ and expectations equal to the stationary values $x_{10}$ and $x_{20}$. The natural boundary conditions for probability density are chosen defined as follows:  $ w(x_{1} ,x_{2} ,t) \longrightarrow 0$   if  $x_{1,2} \longrightarrow + \infty$. The other parameters are specified under the figures. Hereinafter $\theta_{1}=\theta_{2}=\theta$ during simulation.

The appearance of transient bimodality is reflected most clearly on dependencies of the most probable values $x_{1mp}$  and $x_{2mp}$ of intermediate component concentrations on time. In behavior of $x_{1mp}(t)$ a single or a double discontinuity of the first kind is observed. This discontinuity defines the most probable value jump from one maximum $ w(x_{1} ,x_{2} ,t)$  to another. These dependencies are presented in Fig.~\ref{fig1}. They correspond to two values of the bifurcation parameter $B = 6$ (Fig.~\ref{fig1}a) and $ B = 7$ (Fig.~\ref{fig1}b).

Carrying out the simulations with a step 0.01 over $\theta$ we defined the values of $\theta$, at which transient bimodality exists for the values of $B$ above.  Lines of different thickness in  Fig.~\ref{fig1}  correspond to $\theta$ values, at which it first appears and disappears. From the plots presented it follows that for given $B$ this type of solutions exists in a narrow range of $\theta$ values: $0.08<\theta <0.13$ if $B = 6$, and

\begin{figure}[h!]
\includegraphics[width=2.733in,height=1.307in]{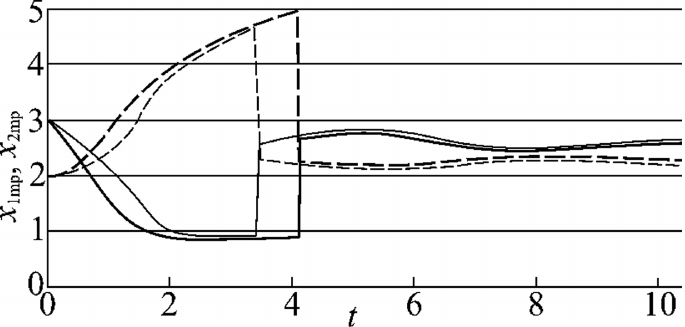}\\
(a)\\
\includegraphics[width=2.752in,height=1.54in]{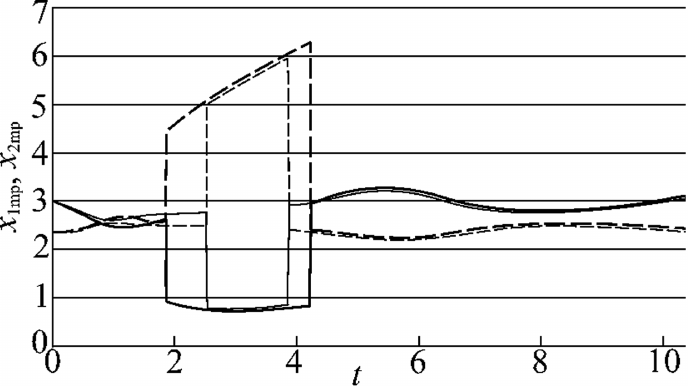}\\
(b)\\
\caption{\label{fig1} The dependencies of most probable values $x_{1mp}$, $x_{2mp}$ of the first and second component concentrations on time for noise intensity values when the transient bimodality appears for the first time (0.5pt lines) and disappears (0.75pt lines) for two values of the bifurcation parameter: (a) $B = 6$, $\theta = 0.09$ - 0.5pt line, $\theta = 0.12$ - 0.75pt line; (b) $B = 7$, $\theta = 0.04$ - 0.5pt line, $\theta = 0.05$ -  0.75pt line. Hereinafter in all figures, the solid lines correspond to the first component, and the dashed lines correspond to the second component.}
\end{figure}

\begin{figure}[h!]
\includegraphics[width=3.102in,height=1.28in]{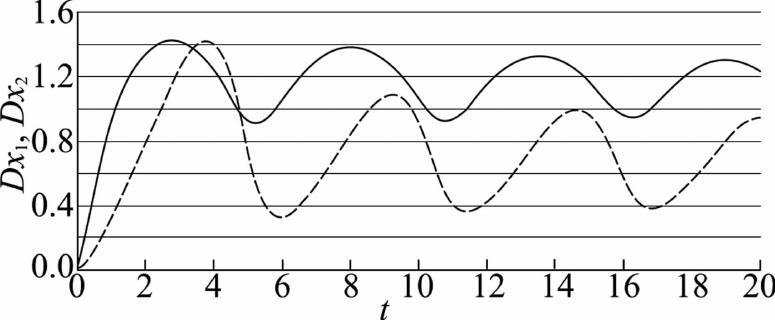}\\
(a)\\
\includegraphics[width=3.047in,height=1.413in]{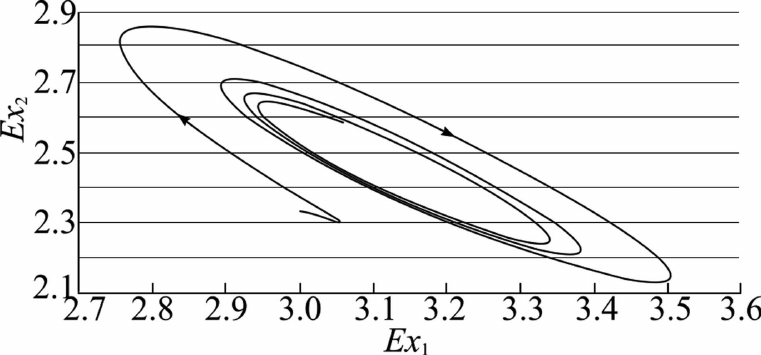}\\
(b)\\
\caption{\label{fig2} The dependencies of statistical characteristics of product concentrations $x_{1,2}$ typical for the regime with the transient bimodality:  (a) variances $Dx_{1}$  and $Dx_{2}$ on time,  (b)  expectations $Ex_{2}(Ex_{1})$. The model parameters are $B = 7$, $\theta = 0.04$.}
\end{figure}

\noindent
$0.03 <\theta <0.06$ if $B = 7$. Thus, at a greater distance from the deterministic bifurcation point the region of existence of the second-type solutions is narrowed. It also follows from the dependencies presented that at a greater distance from the bifurcation point the bimodal regime appears at lower values of $\theta$ and for a given $B$ its duration increases with increasing noise intensity.

Fig.~\ref{fig2}  presents dependencies of variances $Dx_{1}$ and $Dx_{2}$ of component concentrations characteristic for the regime with transient bimodality on time and the curve $Ex_{2}(Ex_{1})$ illustrating the dynamics of the averages. It can be seen from the plots that the variance oscillations are slowly damped. The curve $Ex_{2}(Ex_{1})$ curls up toward the point (3.15, 2.45) presenting a new statistical stationary state which is different from the point of the deterministic stationary state (3.00, 2.33) for the given model (\ref{eq8}) parameters.

When $B = 6$ and $\theta >  0.12$, and $B = 7$ and $\theta > 0.05$  Eq. (\ref{eq9}) demonstrates a third type of solutions. Figs.~\ref{fig3},\ref{fig4} show the characteristic form of probability density evolution for the regime with ``repumping" and the appropriate statistical characteristics of the model (\ref{eq8}). From the figures we can see that under the chosen problem parameters all changes occur almost periodically.

Fig.~\ref{fig3} shows two periods of ``repumping" of the probability density. From Fig.~\ref{fig3} it can be seen that, under the chosen problem parameters the period of probability density repumping is approximately equal to 11 units of model time, and the duration of the bimodal state is about 3 units, i.e. approximately a third of the time the system remains in a disordered state. Thus, the durations of unimodal and bimodal state are comparable in the order of magnitude.

Fig.~\ref{fig4}(a) illustrates the dependencies of variances of concentrations $x_{1}$ and $x_{2}$ on time with increasing noise intensity. We can see from the plots presented that the period of variance change decreases with the growth of noise intensity. If analogous dependencies shown in Fig.~\ref{fig2}(a) at the regime with transient bimodality were of a damped nature, here they take the form that is more like relaxation oscillations. This can be explained by the fact that with increasing noise intensity quadratic nonlinearity, implicit in generalized diffusion coefficients ${C}^{1}_{diff} = {{\theta}_{1}}{{\Phi}_{1,0}} {{x}^{2}_{1}} +  {\theta_{3}}{ \Phi_{3,0}}$,  ${C}^{2}_{diff} ={ \theta_{2} }{\Phi_{2,0}}{ x^{2}_{1}}$  increases.

Changes of the mean and most probable values of concentrations $x_{1}$ and $x_{2}$ in time with increasing noise intensity are presented in Figs.~\ref{fig4}(b,c). Fig.~\ref{fig4}(b) shows that with increasing noise intensity the absolute values of deviations of expectations $Ex_{1,2}(t)$  from the stationary concentrations in deterministic case decrease. Changes of periods of mean and most probable values also decrease. It means that the ``repumping" occurs in increasing frequency.

All changes presented above describe the scenario of disorder origination from ordered state in the system (\ref{eq8}) under the influence of external noise in sequence and in sufficient detail. We can interpret the results presented above as follows. Beginning from some value of external

\begin{figure}
\includegraphics[width=3.133in,height=7.405in]{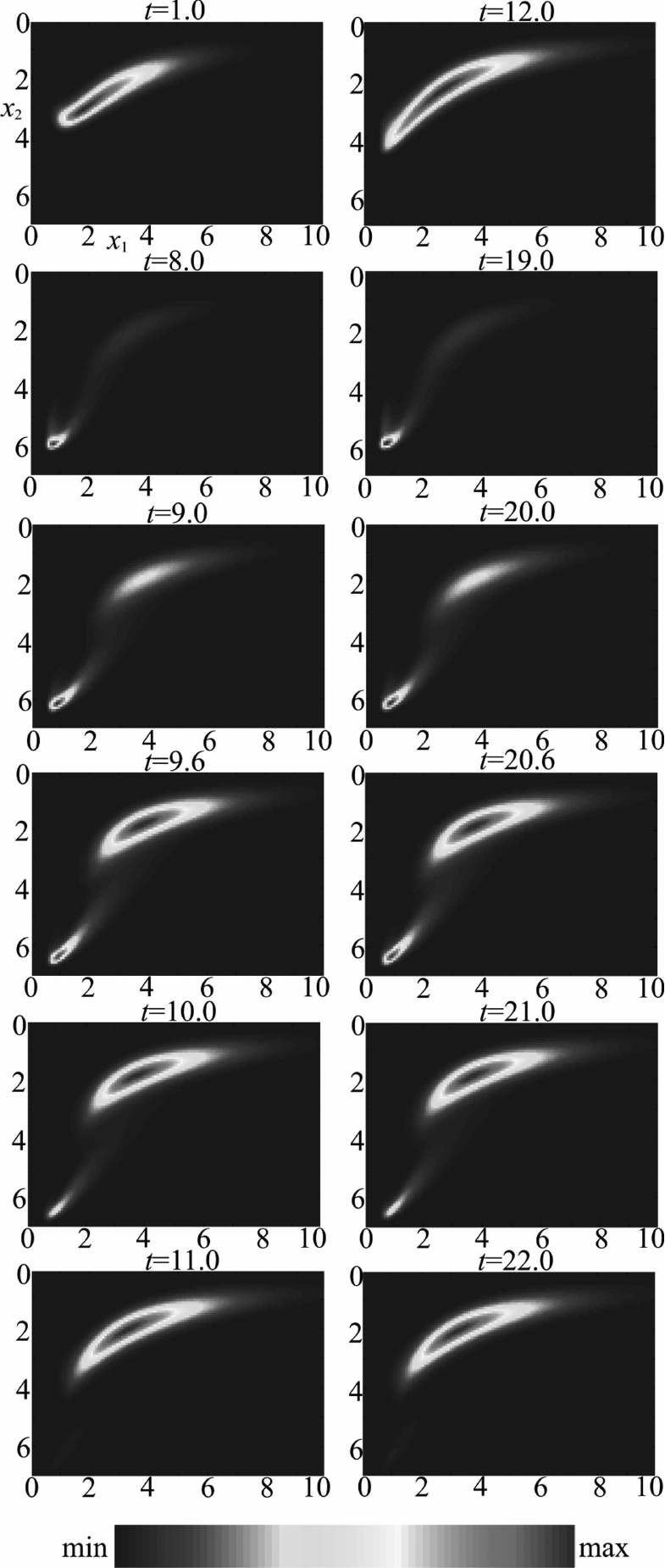}\\
\caption{\label{fig3} The probability density (\ref{eq9}) evolution for the model (\ref{eq8})  corresponding to the regime with ``repumping" through the bimodality. The figure presents two ``repumping" periods. ``Repumping" is observed in the time intervals $ t \in (8,11)$  and $ t \in (19,22)$. The color gradient presented in the figure illustrates the change of  $w(x_{1},x_{2},t)$ from minimum to maximum. The model parameters appropriate to this solution are $B = 7$, $\theta = 0.08$. The dependencies $Dx_{1,2}(t)$, $Ex_{1,2}(t)$, $x_{1,2mp}(t)$ appropriate to this solution are plotted in Fig.~\ref{fig4} as 0.75pt lines (medium thickness lines).}
\end{figure}

\noindent
noise intensity $\theta_{cr1}$ inside the ordered phase disorder originates existing for a finite time, and the higher the noise level, the longer this disorder ``embryo" lives. The farther away from the bifurcation point, the lower the noise that generates it and the narrower the range of noise intensity values at which the system evolves to the ordered, but already a new statistically steady state.

\begin{figure}[h!]
\includegraphics[width=2.638in,height=1.685in]{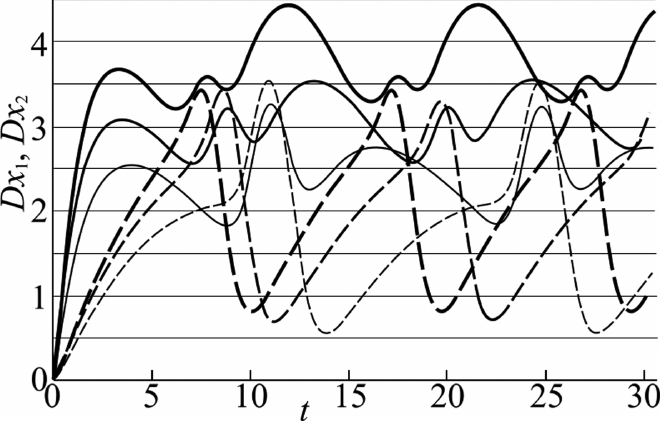}\\
(a)\\
\includegraphics[width=3.122in,height=1.552in]{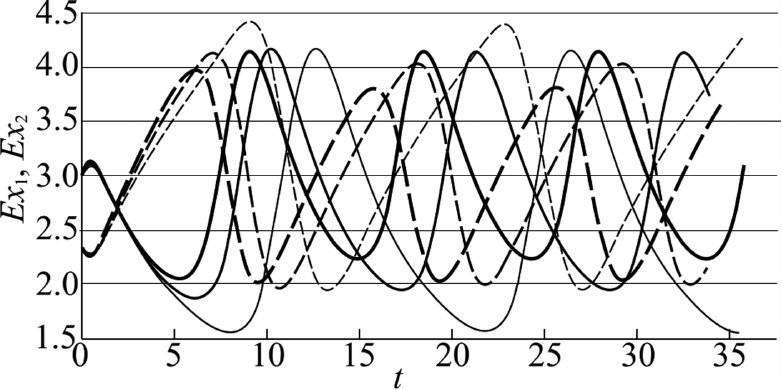}\\
(b)\\
\includegraphics[width=3.012in,height=1.575in]{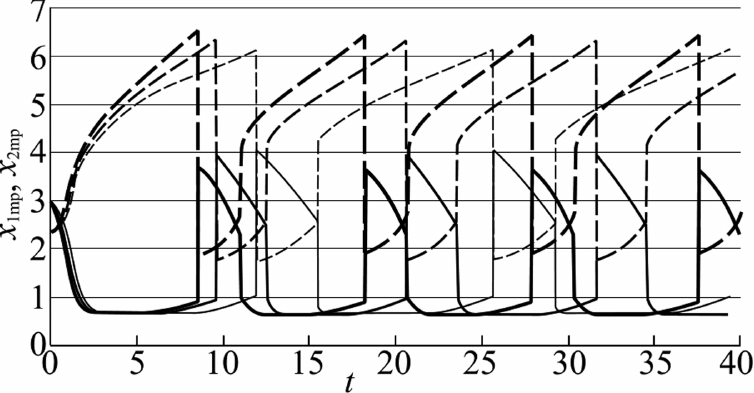}\\
(c)\\
\caption{\label{fig4} The dependencies of statistical characteristics of concentrations $x_{1,2}$ on time with increasing noise intensity in the regime of probability density ``repumping": (a) variances $Dx_{1,2}(t)$, (b) expectations $Ex_{1,2}(t)$, (c) most probable values $x_{1,2mp}(t)$. The model parameters are $B = 7$; $\theta = 0.06$ - 0.5pt line, $\theta = 0.08$ - 0.75pt line, $\theta = 0.10$ - 1pt line.}
\end{figure}

At some second noise intensity value $\theta_{cr2} > \theta_{cr1}$  the ordered and disordered phases begin to alternate repeatedly and almost periodically, and the durations of both phases are comparable in the order of magnitude, i.e. a disordered phase exists on a par with an ordered one. The system dwells in a state of disorder for a significant part of the time. The increasing noise intensity leads to the fact that the ordered and disordered state life times decrease - order and disorder alternate increasingly. In other words, increasing intermittence of ordered and disordered phases is observed, herewith the fraction of time when the system dwells in a disordered state increases.

It is obvious that when the noise exceeds some threshold $\theta_{cr3}$ ($\theta_{cr3} > \theta_{cr2}$) the disorder begins to prevail - the ordered phase is completely destroyed.

Thus, the scenario of the noise induced order-disorder transition consists in the intermittency of the ordered and disordered phases. The intermittency frequency is increased with the growth of the noise intensity.

\section{Simulation}

How does the presence of disorder inside the ordered state affect the process of spatial pattern formation?
Spatial pattern formation is accompanied by increasing variance of functions characterizing the system state. Even in the absence of external noise variance reaches macroscopic values. If patterns are destroyed variance decreases. For the system (\ref{eq8})  it can be seen especially clear for $ Dx_{2}$. From  Fig.~\ref{fig4}(a) we see that variance $ Dx_{2}$  (0.75 pt line for given model parameters) sharply increases to the moments of model time 8 and 19 units. In Fig.~\ref{fig3}  unimodal probability density is appropriate to these moments, i.e. the system is ordered, and, hence, a spatially inhomogeneous pattern of some configuration is formed. In the time intervals (8,11) and (19,22) we see a bimodal probability density in Fig.~\ref{fig3} and sharp decreasing variance $ Dx_{2}$  in Fig.~\ref{fig4}(a). In these time intervals the system is disordered and the pattern formed earlier is destroyed. Thus, in the system evolution we must observe consequent change of spatial pattern configurations.

We carried out a numerical simulation of the system (\ref{eq8}) evolution. Fig.~\ref{fig5} presents the concentration $x_{1}$ distribution at different moments of model time when labyrinth patterns are well formed. A steady stationary state is established in the absence of noise for such model parameters, the pattern formed does not change its configuration in this state. Here we observe the following. Pattern configuration changes at certain time moments: some labyrinth lines merge, others separate or obtain other outlines. Regions, where such changes occurred, are indicated in  Fig.~\ref{fig5}  with circles. Thus, we observe a consequent change of the spatial inhomogeneous configurations, which was predicted above.

\onecolumngrid{

\begin{figure}[h!]
\includegraphics[width=5.162in,height=1.102in]{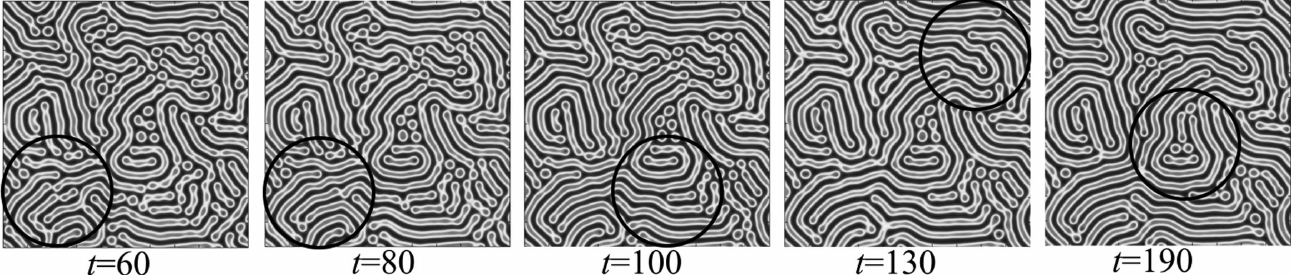}\\
\caption{\label{fig5} Concentration $x_{1}$ evolution in the ``Brusselator" model (\ref{eq8}). Here regions are selected inside which the pattern is changed in comparison with the previous frame. We observe consequent change of the labyrinth configurations. The model parameters are as in Fig.~\ref{fig3}.}
\end{figure}
\begin{figure}[h]
\includegraphics[width=5.153in,height=1.115in]{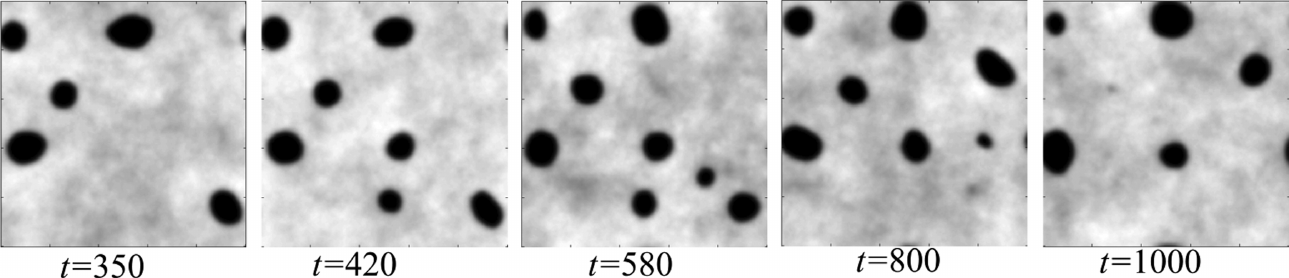}\\
\caption{\label{fig6} Change of the spatial pattern of the ``spots"  configuration under the influence of the noise in the model ``phytoplankton- zooplankton" (according to Ref.\cite{Kurushina55}).}
\end{figure}
}

\twocolumngrid{
Our theory is also confirmed by the results of research of the Turing pattern evolution under the influence of strong noise in a biophysical model ``phytoplankton-zooplankton" described in detail and studied in Ref.\cite{Kurushina55}. Fig.~\ref{fig6} demonstrates consequent change in time of configurations of patterns such as ``spots": some spots disappear, others appear, their size and form change.

Thus, the numerical simulation of the evolution of at least two models confirm qualitatively predicted in the mean-field approximation scenario of noise-induced order-disorder transition.

It should be noted that the ``Brusselator" and ``phytoplankton-zooplankton" models are both a special case of the generalized model (\ref{eq1}), but nonlinear function $g_{i}$  in these models are of different types: in the ``Brusselator" model it is the simplest linear function, whereas in the ``phytoplankton-zooplankton" model it is a fractional rational function with a quadratic nonlinearity in the numerator. However, the observed effect is the same. So, the above scenario of noise-induced order-disorder transition in the model (\ref{eq1}) is apparently universal, at any rate when $n = 2$.

\section{Conclusion}

In our paper we studied numerically consequently and in sufficient detail two types of solutions of  NSCFPE written in MFA for the spatially extended stochastic model ``Brusselator": a solution at which transient bimodality appears, and a solution with multiple alternation of unimodal and bimodal types of probability density distribution. Studying of the probability density dynamics and the model statistical characteristics appropriate to these solutions allowed us to describe the origination and growth of the disorder inside ordered state under the influence of external noise and to determine the scenario of noise-induced order-disorder transition. It is shown that the order-disorder transition is performed through intermittency of ordered and disordered phases, the frequency of intermittency grows with the increasing of noise intensity. Numerical simulation of the evolution of the system under study confirmed the result predicted in MFA: ordered and disordered phases intermittency affects the process of spatial pattern formation as a consequent change of spatial inhomogeneity configurations.

}
\end{document}